# Plasmon Soft Mode in an Organic-inorganic Hybrid Perovskite

Jiyu Tian,[1] Karuna Kara Mishra,[2] E. Zysman-Colman,[1] F. D. Morrison,[1] R. S. Katiyar,[2] and J. F. Scott[1,3]

[1]*EaStCHEM School of Chemistry, University of St. Andrews, St. Andrews, UK KY16 9ST*
[2]*SPECLAB, Dept. of Physics, University of Puerto Rico, San Juan, PR 00931-3343 USA*
[3]*School of Physics, University of St. Andrews, St. Andrews, UK KY19 9SS*

**Abstract**

We report inelastic light scattering from underdamped plasmons in azetidinium lead bromide (AzPbBr$_3$). The plasmons are very strongly temperature dependent and serve as a soft mode for the semiconductor-insulator phase transition near T$_C$ ≈ 150 K, demonstrating a continuous decrease in hole concentration $n_p$(T) by at least a factor of four and implying a nearly tricritical transition. The plasmon frequency and linewidth agree with independent measurements, and the impedance analysis reveals a frequency dependence (modelled by a constant phase element, CPE) that can be identified as due to electron-phonon coupling. The dependence of plasmon frequency upon (T$_C$-T) is analogous to that for magnons in magnetic insulators or soft transverse optical phonons in ferroelectrics and ferroelastics, or for phasons in incommensurately modulated insulators.

**Introduction**

In condensed matter physics or solid-state chemistry we are used to the idea that some boson excitation becomes unstable to signify a phase transition. This has included at least four kinds of bosons: optical phonons (ferroelastic quartz[1] or nearly ferroelectric strontium titanate[2–5]), magnons (antiferromagnets[6] or ferromagnetic insulators), phasons in incommensurately modulated systems[7] and Cooper pairs (superconductivity),[8] Table 1. In the present work we extend this list to show that a semiconductor-insulator transition is triggered by an unstable plasmon ("soft plasmon") in an organic-inorganic hybrid perovskite (OIHP) material.[9] For the past few years there has been a general question as to whether the electrons form localized excitons or coherent plasma waves (plasmons); the plasmon model was verified first by Saba et al.[10]

Here we report these plasmons in azetidinium lead bromide (AzPbBr$_3$)[11] via resonant Raman spectroscopy. The underdamped plasmons shown in Fig. 1 have frequencies between 300-600 cm$^{-1}$, depending upon temperature, and both frequency and linewidth agree with those calculated independently from carrier concentration, $n_p$(T), effective mass, $m^* = ca.$ 0.16 m$_e$, and mobility, $\mu$(T).

The plasmons exhibit a unique and unexpected temperature dependence, with frequency decreasing by >50% as temperature is increased toward T$_C$ ≈ 150 K, suggesting a metal-insulator (or more precisely, semiconductor-insulator) phase transition. Because the phase transition is slightly first-order, it is not routine or reliable to try to extract a temperature exponent, β, from plasmon frequency ω = A(T$_C$-T)$^β$. However, if we merely graph log ω versus log (T$_C$-T), we obtain a good estimate of β = 0.41; and if we instead force T$_C$ to fit independently a peak in the dielectric constant at T ≈ 150 K, we obtain a value β = 0.5, Fig. 2b. So either suggest a mean-field dependence. At this temperature there is a small, presumably secondary effect, on the dielectric properties. The plasmon data are remarkably similar to those in CdS, for which studies were published long ago;[12] this is expected since the effective masses (*ca.* 0.16 $m_e$) and carrier concentrations (a few times 10$^{17}$ cm$^{-3}$) are similar. This recently reported addition to the family of OIHP, AzPbBr$_3$, is a wide-band direct-gap p-type semiconductor ($E_g$ = 2.81 eV*)*, very similar to II-VI's such as ZnTe in many respects.[13–17]

We note parenthetically that boson excitations are characterized by quantum superposition, so that in large number densities they produce classical fields: For ferroelastic soft modes these are strain fields; for ferroelectrics, they are bulk polarizations, *P*; for superconductors, a magnetization-free region (Meissner effect); and for incommensurate insulators, a lattice constant shift (atomic-scale small-wavelength modulation) shift or twist. In the present case there is an electric field, *E*.

The broad Raman modes shown in Figs. 1b and 1c fit the observed plasmon lineshapes to the expected Lorentzians. The peak frequencies are given approximately by

$$\omega_p = \frac{1}{2\pi c}\left[\frac{4\pi n_p e^2}{\varepsilon_\infty m^*}\right]^{\frac{1}{2}} \tag{1.}$$

and the linewidth

$$\Gamma = \frac{e}{2\pi\mu(T)m^*} \tag{2.}$$

The peaks frequencies were determined both manually from the maximum intensity and also by Lorentzian fitting, fig 1b; the latter also provides peakwidth. Using literature values of $m^*$ = 0.16[18–20] and $\varepsilon_\infty$ = 5.76 (calculated from n = 2.4[21,22]), carrier concentrations, $n_p$ of 2.5 × 10$^{16}$ cm$^{-3}$ and mobilities μ *ca.* 200 cm$^2$/V.s were determined at 135 K. The carrier density increases by a factor of four on cooling to 83 K, Fig 1c, whereas the mobility increases to *ca.* 300 cm$^2$/V.s in the same range. In contrast to the data on CdS, there is no strong interference in the phonon region;[14] this is because CdS has a much greater Frohlich interaction between electrons and phonons.

Bulk plasmons occur where the total low-frequency dielectric constant, phonon-part plus plasma, $\varepsilon(0) = 0$. Surface plasmons occur at slightly lower frequencies, nearer the transverse optical phonon branch, where in flat films or interfaces $\varepsilon(0) = -1$, and in small spherical particles where $\varepsilon(0) = -2$.

The plasmons will be underdamped only up to wave vector $q$ given by the reciprocal screening length (Landau damping). For our range of carrier concentrations $n_p(T)$ this is 100-200 nm in real space (average electron separation) or $q < 1 \times 10^7$ cm$^{-1}$. Ceramic samples permit only Raman backscattering geometries, with wave vector transfer

$$q = q_{laser} - \left[-q_{scattered}\right] = 2\times 2\pi n\omega_{laser} = 6.9\times 10^6 \text{cm}^{-1} \tag{3.}$$

So, in our experiments the plasmon linewidths are probably limited by Landau damping. Future experiments at small scattering angles with single crystals should therefore reveal narrower linewidths. Note that this wave vector is only *ca.* 1% of the way to the Brillouin zone boundary, so inelastic neutron scattering is not a useful tool.

The soft plasmon frequencies satisfy

$$\omega(T) = \omega(0)\left[(T_C - T)/T_C\right]^\beta \tag{4.}$$

with $\beta$ estimated in two different ways as *ca.* 0.4 (freely adjustable $T_C$) and 0.5 ($T_C$ constrained to be that of the dielectric anomaly $T_0$), compatible with mean field; the latter is shown in Fig. 2. However, these are only rough estimates, since $T_C$ is unknown and the transition is slightly first-order (note that $T_C$ must be $>T_0$, the actual transition temperature).

**Dielectric properties and CPE**

The dielectric properties of AzPbBr$_3$ show a large dielectric anomaly in the real permittivity at *ca.* 173 K which is associated with a structural transition;[11] at lower temperature, in the range where the plasmons are observed in the Raman data, a clear dielectric loss peak is evident together with a frequency dependence of the real permittivity, Figs. 2a and 2b. These features suggest a contribution of the high frequency plasmons to the low frequency dielectric response. The semiconductor-insulator transition associated with the plasmon response, which is associated with the frequency dispersion, can be fitted to a CPE (constant phase element) which is derived from Jonscher's 'universal' dielectric response.[23,24] This is a circuit element with an complex admittance:

$$Y^*_{CPE}(\omega) = A_0(i\omega)^n \qquad (5.)$$

where $A_0$ is the reciprocal electrical impedance ($|Z^*|$) at 1 rad/s and where $0 \leq n \leq 1$, indicating a range of non-ideal behaviour ranging from $n = 1$ for $Y^*(\omega)$ for a perfect capacitor to $n = 0$ for that of a resistor. Using the relation $Y^*(\omega) = i\omega C_0\varepsilon^*(\omega)$, the complex permittivity $C_0\varepsilon^*(\omega)$ is given by:

$$C_0\varepsilon^*_{CPE}(\omega) = A_0\omega^{n-1} \qquad (6.)$$

This allows the frequency dispersion of the real permittivity to be quantitatively fitted as $\varepsilon' \propto \omega^{1-n}$ (where $n = 1$ indicates an ideal, non-dispersive dielectric response). CPE fits of the real permittivity indicate a minimum $n$ of *ca.* 0.995 (maximum dissipation) in agreement with the dielectric loss data, fig 2c. The use of CPE elements is not unusual in impedance analyses of ferroelectrics[17,25] and ionic conductors,[26] but the origin in each case is physically typically unknown – in many cases it is an ad hoc "fudge factor" yielding an exponent $n$ that can represent multiple deviations from ideality, including electrical inhomogeneity. We find it to be 0.995 here, a value typical of crystalline solids.[17,25,26] This seems to be a rare case in which the physical origin of the CPE is known; it presumably arises from plasmon-phonon coupling, and in particular, through Landau damping into the same final states.

**Role in superconductivity in oxides**

The occurrence of superconductivity in materials such as SrTiO$_3$ remains an important and timely question, with very recent papers in high-impact journals. The temperature of superconductivity is the highest in K/electron of any known material (roughly 0.3 K at 10$^{19}$ cm$^{-3}$), suggesting some pairing mechanism is still missing from models. These may involve soft, very low frequency, underdamped plasmons. In this pairing context Baratoff and Binnig[16] first pointed out the strong coupling between electrons and high-frequency longitudinal optical phonons in SrTiO$_3$. These soft plasmons may also play a role in the dynamics of superconducting interfaces, as in SrTiO$_3$/LaAlO$_3$.[27]

In conclusion, we have shown that a soft plasmon mode can characterize a semiconductor-insulator phase transition. This completes an earlier quartet of soft phonons, soft magnons, soft phasons, and Cooper-pair driven instabilities in ferroelectric/elastics, antiferromagnets, incommensurates, and superconductors; each now can be represented by a boson-like excitation with correspondence as a "classical" field. Further work will involve evaluating other critical exponents, such as isothermal susceptibility divergence gamma or field exponent (usually termed delta). Second, it is important to verify if these soft plasmons are characteristic in other organic inorganic hybrid perovskite (OIHP) materials: Are they generic? If so, can one extend the present measurements to optically illuminated specimens to verify the photo-induced changes in carrier concentration and mobility, both dc (steady-state) and pulsed?

Contributions from each co-author: These samples were synthesised by JT; Raman data were from KKM, RSK, and JFS. All authors contributed to the writing of the manuscript. The plasmon study was organized by JFS.

We thank the Chinese Scholarship Council for Ph.D. Studentship support (to JT). Funding support was from the EPSRC Grants No. EP/P024637/1 and No. EP/P024904/1 and from DoD-AFOSR (Grant #FA9550-16-1-0295).

The authors confirm that they have no financial interests in this work.
Correspondence and requests for materials should be addressed to jfs4@st-andrews.ac.uk.

**Table 1.** Boson-like excitations as order parameters for phase transitions

| Phenomenon | Excitation | Field | References |
|---|---|---|---|
| Ferroelectricity or ferroelasticity | optical phonon (transverse) | Polarization P<br>Strain S | 1-5 |
| Magnetism | Magnon | H (bulk) | 6 |
| Incommensurate modulated | Phason | shift/twist | 7 |
| Superconductivity | Cooper pairs | Meissner effect | 8 |
| Semiconductor-insulator | Plasmon | E | present work |

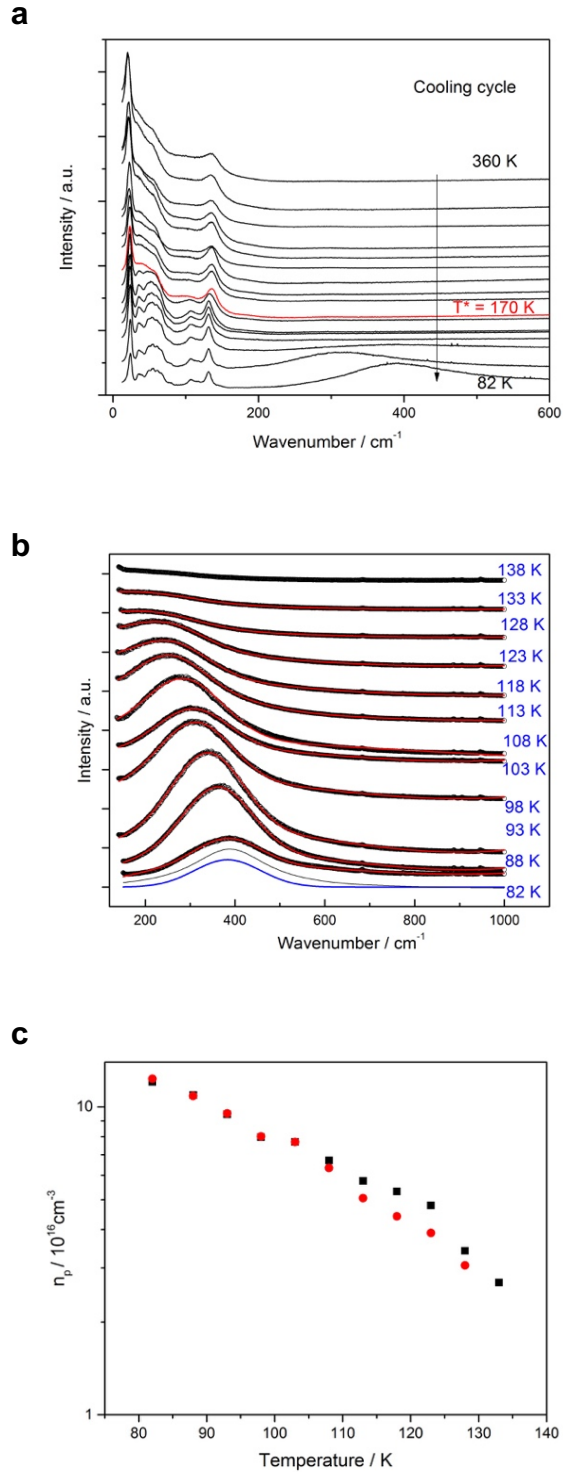

Fig. 1. (a) Raman data of ceramic azetidinium lead bromide, AzPbBr$_3$, at wavelength 514.5 nm and 100 mW, with spectrometer slit width of 200 microns. CCD detector. Data taken on heating showed no noticeable changes from those on cooling (no obvious hysteresis), and the transition temperature marked near T* = 170 K is a known structural transition.; (b) Raman spectra at smaller temperature increments showing evolution of the plasmon peak, with data fitted to Lorentzians. (c) Hole concentration, $n_p$, versus T calculated from the peak position determined both manually (squares) and by Lorentzian fitting (circles).

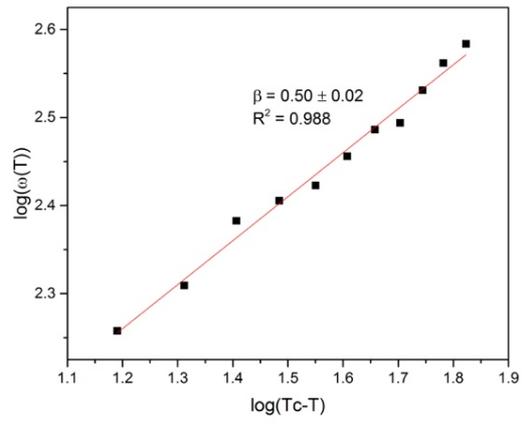

Fig. 2. Plot of ω(T) versus T, constraining $T_C$ (148.5 K) to be that of the onset of the dielectric peak; this gives β = 0.50 exponent in ($T_C$-T).

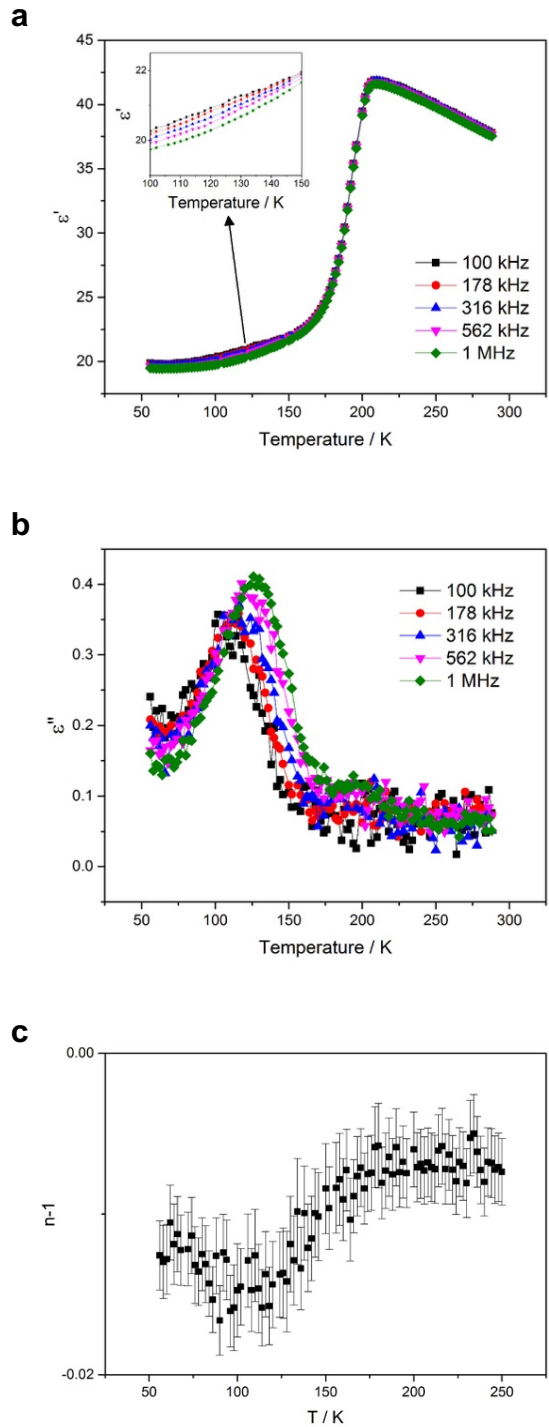

Fig..3. Real (a) and imaginary; (b) permittivity as a function of temperature for azetidinium lead bromide, showing relaxor-like frequency dependence; (c) Fit of the real permittivity to a CPE element (exponent n) indicating the frequency dependence of $\varepsilon' \propto \omega^{n-1}$.